\documentclass[sigconf]{acmart}

\usepackage[english]{babel}
\usepackage{stmaryrd}
\usepackage[utf8]{inputenc}
\usepackage{xspace}
\usepackage{mathtools}
\usepackage[characters]{ltl}
\usepackage{todonotes}
\usepackage{color}
\usepackage{standalone}
\usepackage{natbib}
\usepackage{listings}
\usepackage{amsmath}
\usepackage[capitalize]{cleveref} 
\usepackage{multirow}
\usepackage{rotating} 
\usepackage{adjustbox}
\usepackage{flushend}

\renewcommand{\phi}{\varphi}

\newcommand{\Nat}{\mathbb{N}}
\newcommand{\sseq}[1]{\bar{#1}}  
\newcommand{\subst}[2]{#1\mapsto #2}

\DeclarePairedDelimiter\abs{\lvert}{\rvert}%

\newcommand\wloop{\textbf{loop (*)}}
\newcommand\wforall{\textbf{forall}\xspace}
\newcommand\wchoose{\textbf{choose}\xspace}
\newcommand\wor{\textbf{or}\xspace}
\newcommand\wstmts{\textit{stmts}}
\newcommand\wstmt{\textit{stmt}}
\newcommand\wmay{\textbf{may}\xspace}
\newcommand\wadd{\mathbin{{+}{=}}}
\newcommand\wsub{\mathbin{{-}{=}}}
\newcommand\wpm{\mathbin{{\pm}{=}}}
\newcommand\Oracle{\ensuremath\mathit{Oracle}}

\newcommand\aux{\textit{Choice}}
\newcommand\sem[1]{\ensuremath{\llbracket #1 \rrbracket}}

\newcommand{\blocks}{\mathit{blocks}}

\newcommand\xcomp{\ensuremath{x}\text{-comp}\xspace}
\newcommand\ycomp{\ensuremath{y}\text{-comp}\xspace}
\newcommand\adjy{\ensuremath{\mathit{Adj}}\xspace}
\newcommand\reachy{\ensuremath{\mathit{Reach}}\xspace}

\newcommand\obs{\ensuremath{\mathit{Obs}}\xspace}
\newcommand\afirst{\ensuremath{a_{\mathit{first}}}}
\newcommand\alast{\ensuremath{a_{\mathit{last}}}}

\newcommand\causal{\ensuremath{\mathit{causal}}}
\newcommand\stubborn{\ensuremath{\mathit{stubborn}}}
\newcommand\textcausal{causal\xspace}
\newcommand\Textcausal{Causal\xspace}
\newcommand\textstubborn{stubborn\xspace}
\newcommand\noninter{\ensuremath\mathit{noninterferent}}

\newcommand\lowinputs{\ensuremath\mathit{same\_low\_inputs}}
\newcommand\lowoutputs{\ensuremath\mathit{same\_observations}}
\newcommand\highinputs{\ensuremath\mathit{same\_declassified\_high\_inputs}}
\newcommand\samecfg{\ensuremath\mathit{same\_paths}}

\newcommand\ltl{\textnormal{LTL}\xspace}

\newcommand\fol{\textnormal{FOL}\xspace}
\newcommand\fotl{\textnormal{FOLTL}\xspace}
\newcommand\hfotl{\textnormal{HyperFOLTL}\xspace}
\newcommand\bsfotl{\text{$\exists^*$\fotl}\xspace}
\newcommand\BS{Bernays-Schönfinkel\xspace}

\newcommand{\true}{\textit{true}}
\newcommand\false{\textit{false}}
\renewcommand\implies{\rightarrow}

\newcommand\equ{\leftrightarrow}

\newcommand{\vars}{\mathcal{V}}
\newcommand{\fv}{\mathit{fv}}

\newcommand{\sig}{\Sigma}  
\newcommand{\sorts}{S}  
\newcommand{\rels}{\mathcal{R}}
\newcommand{\hirels}{\mathcal{R}_{\mathit{high}}}
\newcommand{\lirels}{\mathcal{R}_{\mathit{low}}}

\newcommand{\cfgrels}{\mathcal{R}_{\mathit{cfg}}}
\newcommand{\wfrels}{\mathcal{R}_{\mathit{wf}}}
\newcommand{\csts}{\mathcal{C}}
\newcommand{\ar}{\mathit{ar}}
\newcommand{\valuation}{\nu}
\newcommand{\ofType}[1]{{:}\,{#1}}

\newcommand{\struct}{\ensuremath{\mathcal{S}}}  

\newcommand{\tstruct}{\bar{\struct}} 

\newcommand{\fcfg}{\ensuremath{\mathtt{cfg}}}
\newcommand{\fsanity}{\ensuremath{\mathtt{sanity}}}
\newcommand{\fblocks}{\ensuremath{\mathtt{exec}}}
\newcommand{\finitial}{\ensuremath{\mathtt{init}}}

\newcommand{\fwf}{\ensuremath{\mathtt{wf}}}

\newcommand{\lastnode}{n_{\mathit{end}}}

\newcommand\globally{\LTLglobally}
\newcommand\eventually{\LTLfinally}
\newcommand\nextstep{\LTLnext}
\newcommand\until{\LTLuntil}
\newcommand\release{\LTLrelease}
\newcommand\wuntil{\LTLweakuntil}

\newcommand\Conf{\textit{Conf}} 
\newcommand\Assign{\textit{Assign}} 
\newcommand\Read{\textit{Read}} 
\newcommand\Review{\textit{Review}} 

\newcommand\ttrue{\ensuremath{\mathit{true}}\xspace}

\newcommand\aalta{\textsc{Aalta}\xspace}
\newcommand\niwo{\textsc{NIWO}\xspace}

\lstdefinelanguage{workflows}{
  morekeywords={loop, forall, may},
  columns=fullflexible,
  sensitive=true,
  commentstyle = \itshape, 
  morecomment={[l]//},
  mathescape=true,
  basicstyle=\small,
  identifierstyle={\ttfamily},
  literate={<-}{$\leftarrow\ $}{2} {->}{$\rightarrow\ $}{2} {:=}{$\leftarrow\ $}{2}
}

\copyrightyear{2017}
\acmYear{2017}
\setcopyright{acmlicensed}
\acmConference{CCS '17}{October 30-November 3, 2017}{Dallas, TX, USA}
\acmPrice{15.00}
\acmDOI{10.1145/3133956.3134080}
\acmISBN{978-1-4503-4946-8/17/10}

\newif\ifinv
\invtrue
\invfalse

\begin{document}

\title{Verifying Security Policies in Multi-agent Workflows with Loops} 
\author{Bernd Finkbeiner}
\email{finkbeiner@cs.uni-saarland.de}
\affiliation{CISPA, Saarland University}

\author{Christian Müller}
\email{christian.mueller@in.tum.de}
\affiliation{Technische Universität München}

\author{Helmut Seidl}
\email{seidl@in.tum.de}
\affiliation{Technische Universität München}

\author{Eugen Z\u alinescu}
\email{eugen.zalinescu@in.tum.de}
\affiliation{Technische Universität München}

\begin{abstract}
  We consider the automatic verification of information flow security policies
of web-based workflows, such as conference submission systems like EasyChair. Our workflow description language allows for loops, non-deterministic choice,
and an unbounded number of participating agents. The information flow
policies are specified in a temporal logic for hyperproperties.
We show that the verification problem can be reduced to the
satisfiability of a formula of first-order linear-time temporal logic, and provide decidability results
for relevant classes of workflows and specifications.
We report on experimental results obtained with an implementation of our approach on a series of benchmarks.

\end{abstract}




\maketitle

\section{Introduction}\label{sec:intro}

Web-based workflow systems often have critical information flow
policies.  For example, in a conference management system like
EasyChair, the information about a certain paper must be kept secret
from all program committee (PC) members who have declared a conflict of
interest for the paper until the acceptance notifications are released by the PC chair.

Verification techniques for workflows (cf.~\cite{BauereissH14,Kanav2014,ATVA16})
typically build on classic notions of secrecy such as
non-interference~\cite{GoguenMeseguer82}.  The particular challenge
with verifying web-based workflow systems is that here is no fixed set
of agents participating in the workflow. Clearly, we would not like to
reason about the correctness of a conference management system for
every concrete installation for a particular conference, a particular
program committee and a particular set of submissions and reports. Instead,
we would like to prove a given system once for all --- for any possible
instantiation and any number of PC members, submitted papers and reports.

We present such a verification approach based on the temporal logic
HyperLTL~\cite{clarkson2014temporal}. HyperLTL is a general specification language for temporal
hyperproperties, which include common information flow policies
like non-interference, and time- and data-dependent declassification.
HyperLTL can also express assumptions on
the behavior of the agents such as \emph{causality}, i.e. that an
agent can only reveal information that was received by the agent at a
\emph{previous} point in time. This is important in order to analyze
chains of information flows, where a piece of information is
transmitted via two or more communications, i.e. where agent A learns about a
secret known to agent B, even though B never communicates with A directly; instead,
B talks to a third agent C, and, subsequently, C talks to A.

HyperLTL-based workflow verification has been considered before,
but only for the restricted case of \emph{loop-free} workflows~\cite{ATVA16}.
Such workflows consist of a fixed finite sequence of steps.
Although an arbitrary number of agents may participate in each step,
the workflow thus only allows a fixed number of interactions
between the agents. This is not realistic: to accurately model,
for example, the repeated commenting on papers and reviews during
the discussion phase of a conference management system, one needs
a loop in the workflow.

We present an automatic verification technique for
workflows \emph{with loops}.
The general outline of our approach is as follows: We specify the
operational semantics of the workflow language in many-sorted
first-order linear-time temporal logic (\fotl).  The desired
information-flow policy and the assumptions on the agents are
expressed in first-order HyperLTL (\hfotl). Combining the two
specifications, the existence of a violation of the policy reduces to
the satisfiability of a \hfotl formula.

We identify an expressive fragment of many-sorted \hfotl, for which
satisfiability is decidable.  The fragment subsumes the previously
known decidable fragments of \fotl~\cite{kuperberg2016finite} and of
HyperLTL~\cite{finkbeiner2016deciding}. It also generalizes the \BS
fragment of first-order logic~\cite{AbadiRS10}. Of particular
practical value is that our logic is \emph{many-sorted}, i.e. we
distinguish different groups of agents such as authors and
program committee members. This allows us to place different
assumptions on different groups; it also improves the performance
of our decision procedure, because the different sorts are kept
separate.

We identify a natural class of workflows, which we call \emph{non-omitting} workflows,
where the encoding of the verification problem is in the decidable fragment of \hfotl. We thus obtain a decision procedure for non-omitting workflows.
This decidability result in fact turns out to be optimal in the sense that for workflows
outside the class, non-interference becomes undecidable.
The decidable fragment is also sufficiently expressive to specify common information-flow policies like non-interference. In terms of agent assumptions, we show that the fragment is sufficiently
expressive to handle strong assumptions like \emph{stubbornness}, meaning that an agent does not reveal any information, for arbitrary sets of agents, and weaker assumptions, like \emph{causality}, for a
fixed finite set of agents. This means that we can decide whether a given number of agents can
\emph{conspire} to cause a leak, assuming that all other agents do not reveal any information.
Again, our decidability result is optimal in the sense that
the verification problem for unbounded sets of causal agents turns out to be undecidable: it is impossible
to decide whether an unbounded number of agents can conspire to reveal a secret.

We report on experimental results based on an implementation of our approach in the tool \niwo. For example, \niwo has found an attack on a simple conference management system, where two program committee members conspire to leak a secret.

\section{Preliminaries}
\label{sec:prelim}

Given a sequence $\bar\sigma$, we let $\sigma_n$ denote its {$n$-th
  element}, and $\bar\sigma[n,\infty]$ denote its subsequence from $n$
to $\infty$, i.e.
$\bar\sigma[n,\infty] := \sigma_n\sigma_{n+1}\ldots$, assuming
$\bar{\sigma}$ is infinite.
We sometimes abuse notation and use set notation over sequences. For
instance, $|\bar\sigma|$ denotes the length of~$\bar\sigma$.

\subsection{First-Order LTL (FOLTL)}

A \emph{signature} $\sig=(\sorts,\csts,\rels,\ar)$ consists of a
non-empty and finite set of sorts, finite and disjoint sets~$\csts$
and~$\rels$ of constant and relation (or predicate) symbols, and arity
function $\ar:\csts\cup\rels\rightarrow \sorts^*$, with $|\ar(c)|=1$
for any $c\in\csts$, where $\sorts^*$ denotes the set of finite
sequences of sorts.
For each sort~$s$, we let $\vars_s$ be a countably infinite set of
variables. 
We let $\vars := \bigcup_{s\in\sorts}\vars_s$.

\fotl \emph{formulas} over the signature $\sig=(\sorts,\csts,\rels,\ar)$ are
given by the grammar
\begin{displaymath} 
  \phi ::=  t = t' \mid  
  R(t_1,\ldots,t_{k}) \mid 
  \neg \phi \mid \phi \lor \phi \mid \exists x\ofType{s}.\,\phi \mid 
  \nextstep \phi \mid \phi \until \phi\, 
\end{displaymath}
where $t$, $t'$, and the $t_i$s range over $\vars\cup\csts$, $R$
ranges over $\rels$, $s$ ranges over~$\sorts$, and $x$ ranges
over~$\vars_s$.
The symbols~$\nextstep$ and $\until$ denote the usual Next and Until
LTL operators.
As syntactic sugar, we use standard Boolean connectives such as
$\land, \to, \equ$, the universal quantifier~$\forall x$, and the
derived temporal operators $\eventually$ ({Eventually}) with
$\eventually\phi := \true \until \phi$, $\globally$ ({Globally}) with
$\globally\phi := \neg\eventually\neg\phi$, $\wuntil$ ({Weak Until})
with $\phi \wuntil \psi := (\phi \until \psi) \lor \globally \phi$,
and $\release$ (Release) with
$\phi\release \psi := \neg (\neg \phi \until \neg \psi)$, where
$\true := (c=c)$ for some $c\in\csts$.

We only consider well-sorted formulas. We omit their definition, which
is as expected; for instance, equality is only allowed over terms of
the same sort.
We may drop the sort in $\forall x\ofType{s}.\,\phi$ when it is
irrelevant or clear from the context and simply write
$\forall x.\,\phi$.

We will sometimes consider that formulas are in \emph{negation normal
  form}, which is obtained by pushing negation inside until it appears
only in front of atomic formulas. When considering this form, the
operators $\land$, $\forall$, and $\release$ are seen as primitives,
instead of derived ones. 
A formula is in \emph{prenex normal form} if it is written as a
sequence of quantifiers followed by a quantifier-free part.

To omit parentheses, we assume that Boolean connectives bind stronger
than temporal connectives, and unary connectives bind stronger than
binary ones, except for the quantifiers, which bind weaker than
Boolean ones and stronger than temporal ones.

The set of \emph{free variables} of a formula~$\phi$, that is, those
that are not in the scope of some quantifier in $\phi$, is denoted
by~$\fv(\phi)$.
%
A formula without free variables is called \emph{closed} or \emph{ground}. 
For a term $t\in\vars\cup\csts$, we let $\fv(t) := \{t\}$ if
$t\in\vars$ and $\fv(t) := \emptyset$ otherwise.

\smallskip

A \emph{structure}~$\struct$ over the signature
$\sig=(\sorts,\csts,\rels,\ar)$ consists of a $\sorts$-indexed family
of (finite or infinite) \emph{universes} $U_s\neq\emptyset$ and
\emph{interpretations} $R^\struct \in U_{s_1}\times \dots U_{s_k}$,
for each ${R\in \csts \cup \rels}$ of sort $(s_1,\dots,s_k)$. We let
$U := \bigcup_{s\in\sorts}U_s$.
A \emph{temporal structure} over~$\sig$ is a sequence
$\tstruct=(\struct_0,\struct_1,\dots)$ of structures over $\sig$ such
that all structures $\struct_i$, with $i\ge 0$, have the same universe
family, denoted $(U_s)_{s\in\sorts}$, and rigid constant
interpretations, i.e.~$c^{\struct_i} = c^{\struct_{0}}$, for all
$c\in\csts$ and $i>0$.

Given a structure, a \emph{valuation} is a mapping
$\valuation: \vars \to U$ with $x$ and $\nu(x)$ of the same sort for
any $x\in\vars$.
For a valuation $\nu$ and tuples $\sseq{x} = (x_1,\ldots,x_n)$ and
$\sseq{d} = (d_1,\ldots,d_n)$, where $x_i\in\vars_{s}$ and
$d_i\in U_{s}$ for some sort~$s$, for each $i$, we write
$\nu[\subst{\sseq{x}}{\sseq{d}}]$ for the valuation that maps each
$x_i$ to~$d_i$ and leaves the other variables' valuation unaltered.
By $\nu(\bar{x})$ we denote the tuple $(\nu(x_1),\dots,\nu(x_n))$.
We extend this notation by applying a valuation $\nu$ also to
constant symbols $c\in\csts$, with ${\nu(c) = c^{\struct}}$.

Let $\tstruct$ be a temporal structure over the signature $\sig$, with
$\tstruct=(\struct_0,\struct_1,\dots)$, $\phi$ a formula over
$\sig$, and $\nu$ a valuation.
We define the relation $\tstruct,\nu\models\phi$ inductively as
follows:
\begin{displaymath}
  \begin{array}{@{}l@{\quad}l@{\quad}l@{}}
    \tstruct,\nu \models t = t' 
    & \mbox{iff} &
    \nu(t) = \nu(t')
    \\
    \tstruct,\nu \models R(\bar{t}) 
    & \mbox{iff} &
    \nu(\bar{t}) \in R^{\struct_0}
    \\
    \tstruct,\nu \models \neg \psi
    & \mbox{iff} &
    \tstruct,\nu \not \models \psi
    \\
    \tstruct,\nu \models \psi \lor \psi'
    & \mbox{iff} &
    \tstruct,\nu \models \psi
    \mbox{ or }
    \tstruct,\nu \models \psi'
    \\
    \tstruct,\nu \models \exists x.\,\psi
    & \mbox{iff} &
    \tstruct,\nu[x\mapsto d] \models \psi
    \mbox{, for some $d\in U$}
    \\
    \tstruct,\nu \models \nextstep \psi
    & \mbox{iff} &
    \tstruct[1,\infty],\nu \models \psi
    \\
    \tstruct,\nu \models \psi \until \psi'
    & \mbox{iff} &
    \mbox{for some }j \geq 0,\ 
    \tstruct[j,\infty],\nu \models \psi', \mbox{and }
    \\ & &
    \quad \tstruct[k,\infty],\nu \models \psi,\ 
    \mbox{for all $k$ with $0\leq k < j$}
  \end{array}
\end{displaymath}

A \fotl formula $\phi$ is said to be \emph{satisfiable} iff there exists a temporal structure $\tstruct$ and a valuation $\nu$ s.t.~$\tstruct,\nu \models \phi$.
It is said to be \emph{finitely satisfiable} iff there exists a temporal structure $\tstruct$ over a finite universe $U$ and a valuation $\nu$ s.t.~$\tstruct,\nu \models \phi$.

We note that unsorted \fotl can be seen as sorted \fotl with just one
sort.

\subsection{FOLTL Decidability}

Since \fotl subsumes First-Order Logic (\fol), \fotl is also
undecidable.  In this paper, we consider formulas of a \BS-like
fragment\footnote{The Bernays-Schönfinkel-Ramsey fragment, also
  called ``effectively propositional'', is one of the first identified
  decidable fragments of \fol~\cite{BorgerGG1997}. It consists of
  those \fol formulas in prenex normal form having the
  $\exists^*\forall^*$ quantifier prefix.} of \fotl, which we name
\emph{\bsfotl}.
To define this fragment we will consider the projection of a sorted
\fotl formula on a sort~$s$, defined as the \fotl formula obtained by
removing all quantifications and terms of a sort different from~$s$.
We refer to~\cite[Definition~17]{AbadiRS10} 
for the formal definition of the projection, and here we only
illustrate it with an example.
Given the formula
$\exists x\ofType{A}.\, \forall y\ofType{B}.\, \exists z\ofType{A}.\,
\neg(x=z) \land P(x,y) \land \globally Q(y,z)$, its projection on the
sorts~$A$ and~$B$ are the formulas
$\exists x.\, \exists z.\, {\neg(x=z)} \land P_2(x) \land \globally
Q_1(z)$ and $\forall y.\, P_1(y) \land \globally Q_2(y)$, respectively.

The \bsfotl fragment of sorted \fotl consists of those closed formulas~$\phi$
 in negation normal form such that, for each sort~$s$,
the projection of~$\phi$ on~$s$ is a formula of the form
\[
    \exists x_1,\ldots,x_k.\, \phi'_s
\]
with $k\geq 0$ and $\phi'_s$ a \fotl formula containing no existential quantifiers.
This definition extends the definition of the \BS-like fragments
in~\cite{AbadiRS10,NelsonDFK12} from \fol to \fotl,\footnote{The
  decidable \fol fragments in~\cite{AbadiRS10,NelsonDFK12} are larger
  than the projection of the \bsfotl fragment to sorted \fol, as they
  also consider function symbols.} and of the \BS-like fragment
in~\cite{kuperberg2016finite} from unsorted \fotl to sorted \fotl.
Note that the previous sample formula is in \bsfotl.

We can try to put an arbitrary \fotl formula in the mentioned form by using the standard transformations that put a \fol formula into prenex normal form, as well as the following equivalences to move existential quantifiers outside of temporal operators:
\[
\phi \until \exists x. \psi \equiv \exists x.\, (\phi \until \psi)
\quad \text{and} \quad
(\exists x. \psi) \release \phi \equiv \exists x.\, (\psi \release \phi),
\]
assuming that $x$ does not occur free in $\phi$.
Note that in particular we have that $\eventually \exists x. \phi \equiv \exists x. \eventually \phi$.
However, the previous equivalences cannot be generalized. For instance, existential quantifiers cannot, in general, be moved over the $\globally$ operator.
Intuitively, $\globally \exists x. \phi$ means that ``for all time points~$t$, there exists an~$x$ such that $\phi$ holds at~$t$''. Thus, in contrast to~\fol, not all \fotl formulas can be put in prenex normal form.

\begin{theorem}[\bsfotl Decidability]\label{thm:decidability} \ 

  \begin{enumerate}
  \item Checking satisfiability of a formula in \bsfotl is equivalent to checking \emph{finite} satisfiability of the same formula.
  \item \bsfotl is decidable.
  \end{enumerate}
\end{theorem}
\begin{proof}
  This proof follows the reasoning for decidability of the
  Bernays-Schönfinkel-Ramsey fragment of~\fol, see
  e.g.~\cite{BorgerGG1997}.

  Consider a closed formula $\phi$ in \bsfotl.  The formula $\phi$ has
  the form $Q_1 x_1\ofType{s_1} \ldots Q_k x_k\ofType{s_k}.\, \psi$,
  where $k\geq 0$, $Q_1,\dots,Q_k$ is a sequence of quantifiers, and
  $\psi$ is an \fotl formula in negation normal form containing no
  existential quantifiers.
  We group the sequence $Q_1,\dots,Q_k$ of quantifiers into maximal
  subsequences of the form $\exists^*\forall^*$. We let $n$ be the
  number of such subsequences, and let $\psi_i$ be obtained from
  $\phi$ be removing the first $i$ groups of quantifiers, for
  $0\leq i\leq n$. Note that $\phi_0=\phi$ and $\phi_n=\psi$.

  We iteratively transform the formula $\phi_0$ into the formulas
  $\psi_1$ to~$\psi_n$.
  We also build the sets $D^i_s$ of constant symbols of sort~$s$, for
  each sort~$s$ and each $i$ with $1\leq i\leq n$.
  Consider step~$i$, with $1\leq i\leq n$. For each sort~$s$, we pick
  a set $C^i_s$ of constant symbols whose cardinality is given by the
  number of existential quantifiers over the sort~$s$ in the $i$-th
  subsequence, and such that $C^i_s\cap C^j_s=\emptyset$ for any
  $0<j<i$. We let $D^i_s=C^i_s\cup D^{i-1}_s$, where $D^0_s$ is a
  singleton containing some constant of sort~$s$.
  For each sort $s$, and each variable~$x$ of sort $s$ bound by an
  existential quantifier from the $i$-th group, we remove the
  existential quantifier and we instantiate $x$ in $\psi_i$ by a
  corresponding constant from $C^i_s$. In this way, all existentially
  quantified variables in $\psi_i$ are instantiated. Next, starting
  from the top-most universal quantifier in $\psi_i$, we iteratively
  replace every subformula of the form $\forall y\ofType{s}.\,\alpha$
  by the finite conjunction over elements of $D^i_s$, namely,
  $\bigwedge_{d\in D^i_s}\alpha[y\mapsto d]$. Let $\psi_{i+1}$ be the
  formula obtained in this manner. 
  Finally, we replace all subformulas of the form
  $\forall y\ofType{s}.\,\alpha$ in $\psi_n$ (recall that $\psi$ may
  have universal quantifiers) by the finite conjunction over elements
  of~$D^n_s$, as above.
  Let $\psi'$ be the formula obtained in this manner.
  It is easy to see that $\phi$ is satisfiable iff $\psi'$ is
  satisfiable. Furthermore, it is also clear that $\psi'$ is
  satisfiable iff it is finitely satisfiable, as we can pick
  $U_s=D^n_s$ as the (Herbrand) universes. Note that by construction
  the universe~$U_s$ is non-empty even when there is no existential
  quantifier over the sort~$s$; in this case $C^i_s=\emptyset$, for
  all $i$ with $1\leq i\leq n$.
  This ends the proof of the first statement of the theorem.

  For the second statement of the theorem, note that $\psi'$ contains
  neither quantifiers, nor variables, and thus its atoms are ground.
  We transform $\psi'$ into an LTL formula by taking the disjunction
  over all combinations of equivalence relations over $D^n_s$ (for
  each sort~$s$) of the formulas obtained by replacing each predicate
  $R(d_1,\dots,d_\ell)$ in $\psi'$ with the atomic propositions
  $R_{(d'_1,\dots,d'_\ell)}$, where $d'_i$ is the representative of
  the equivalence class to which $d_i$ belongs. Furthermore,
  equalities $a=b$ are replaced by $\true$ if $a$ and $b$ are in the
  same equivalence class and by $\false$ otherwise.
  Clearly, the thus obtained formula is equi-satisfiable with $\psi'$.
  We can now conclude by noting that LTL satisfiability is decidable,
  see e.g.~\cite{Sistla85}.
\end{proof}

We also note that there are very few decidability results concerning
\fotl. Besides the \bsfotl fragment, the only other decidable fragment
we are aware of is the monodic fragment~\cite{HodkinsonWZ00}, which
requires that temporal subformulas have at most one free variable. As
will become clear in the next section, this restriction is too strong
for our purposes: we cannot encode workflows by \fotl formulas in this
fragment.

\section{Workflows}
\label{sec:workflows}

In this section, we will define a language of \emph{workflows}.
Our definition of workflows extends the definition of workflows in~\cite{ATVA16}
with \emph{loops} and \emph{nondeterministic choice}. 

Workflows are used to model the interactions of multiple agents with a system.
Agent interactions are recorded by relations.
Updates of these relations are organized into \emph{blocks}. 
These describe operations that a subset of the agents can choose to execute to change the relation contents.
The most basic construct in the description of workflows is a parameterized guarded
update operation to some relation. Such an update is meant to simultaneously be executed
for all tuples satisfying the given guard. Some of these updates may also be \emph{optional},
i.e. may also be omitted for some of the tuples that satisfy the guard.

\begin{figure}
\[
\begin{array}{l@{\quad}l}
& \textit{\% PC members may declare conflicts}\\
(b_1) & \wforall\ x\ofType{A}, p\ofType{P}\ \wmay\ldotp \ttrue \to \Conf \wadd (x,p)\\
& \textit{\% PC members are assigned to papers}\\
(b_2) & \wforall\ x\ofType{A}, p\ofType{P}\ \wmay\ldotp \neg \Conf(x,p) \to \Assign \wadd(x,p)\\
& \textit{\% PC members write reviews for papers}\\
(b_3) & \wforall\ x\ofType{A}, p\ofType{P}, r\ofType{R}\ldotp
  \\&\quad \Assign(x,p) \land \Oracle(x,p,r) \to \Review\wadd (x,p,r)\\
& \textit{\% PC members discuss about the papers}\\
& \wloop\ \{\\
& \quad\textit{\% PC members read all other reviews}\\
(b_4) & \quad\wforall\ x\ofType{A}, y\ofType{A}, p\ofType{P}, r\ofType{R}\ldotp
  \\&\quad\quad \Assign(x,p) \land \Review (y,p,r) \to \Read \wadd (x,y,p,r)\\
& \quad\textit{\% PC members can rethink their reviews}\\
(b_5) & \quad\wforall\ x\ofType{A}, p\ofType{P}, r\ofType{R}\ \wmay \ldotp
  \\&\quad\quad \Assign(x,p) \land \Oracle (x,p,r) \to \Review \wadd (x,p,r)\ \}
\end{array}
\]
\caption{EasyChair-like workflow.}
\label{fig:easychair}
\end{figure}

\begin{example}
\label[example]{ex:easychair}
The workflow in \cref{fig:easychair} models the paper reviewing and review updating of EasyChair.
In this workflow, all PC members are agents.
In a first step, they can declare that they have a conflict of interest with some of the papers.
Then, papers are assigned to reviewers as long as they have not declared a conflict with the respective papers.
Reviewers are then required to write an initial review of their assigned papers.
Afterwards, the discussion phase starts. Here, all reviewers of a paper are shown all the reviews other people wrote for the same paper.
They can then alter their review based on the information they have seen.
This discussion phase continues for multiple turns until the PC chair ends the phase.
\end{example}

\subsection{Workflow Language}

Workflows are defined over signatures $\sig=(\sorts,\csts,\rels,\ar)$, with
$\rels = \wfrels \uplus \hirels$. Symbols in $\wfrels$ denote
\emph{workflow relations}, which are updatable. Symbols in $\hirels$
denote non-updatable relations that contain \emph{high input}
(i.e.~input containing potentially confidential data) to the
workflow. 
%
%
For instance, in \cref{ex:easychair} we have
$\wfrels=\{\Conf,\Assign,\Review,\Read\}$ and
$\hirels=\{\Oracle\}$. 
 
Workflows~$w$ over a signature $\sig=(\sorts,\csts,\rels,\ar)$ are defined by the grammar given in Figure~\ref{fig:workflows},
\begin{figure}
\[
\begin{array}{lrl@{\quad}l}
w &~::=~&  \textit{block} \mid \textit{block w} & \\
& & \hspace*{-0.5em} \mid \wloop\ \{\textit{w}\,\} \mid \wchoose\ \textit{w}\ \wor\ \textit{w} 
& 
\\
\textit{block} &~::=~&  \wforall\ x_1\ofType{s_1}, \ldots, x_k\ofType{s_k}.\ \wstmts \\
        & & \hspace*{-0.5em} \mid \wforall\ x_1\ofType{s_1}, \ldots, x_k\ofType{s_k}.\ \wmay\ \wstmts
        & 
\\
\wstmts &~::=~& \wstmt \mid \wstmt; \wstmts & 
\\
\wstmt &~::=~& \theta \to R \wadd (u_1, \dots, u_n)\\
  & & \hspace*{-0.5em} \mid \theta \to R \wsub (u_1, \ldots, u_n) & 
\end{array}
\]
\caption{Definition of Workflows}\label{fig:workflows}
\end{figure}
where $s_1,\dots,s_k$ with $k\geq 0$ range over sorts in~$\sorts$,
$x_i$ ranges over~variables in $\mathcal{V}_{s_i}$ for each~$i$,
$u_1,\dots,u_n$ with $n\geq 0$ range over terms in~$\vars\cup\csts$,
$R$ ranges over predicate symbols in~$\rels$, and $\theta$ ranges over
first-order formulas over the signature~$\sig$.
For a statement~$\theta \to R \wpm \bar{u}$, we require that
$R\in\wfrels$, $\abs{\bar u} = |\ar(R)|$, and
$\fv(\theta)\cup\fv(\bar{u})\subseteq\bar{x}$, where $\bar{x}$ is the
sequence of variables appearing in the \wforall construct of the block
that contains the statement.
We only consider well-sorted workflows.

\subsection{Semantics}\label{sec:semantics}

We will now give the semantics of workflows directly using \fotl.

\paragraph{\bf Formalization of the control flow.}

Given a workflow~$w$, we consider its control flow graph (CFG), defined as expected.
We add a node~$\lastnode$ to the CFG with only a single, looping,
outgoing edge, encoding a fictitious new last block that can be
reached after the original last block of the workflow. This new
node is used to encode a finite (terminated) execution of the workflow
by an infinite trace, where the last workflow state is stuttered.
Any infinite path through the graph represents thus an execution of the workflow.

All edges are labelled with blocks. 
Note that for each block there is a unique edge that is labeled with
that block. Edges not corresponding to a workflow block are labeled
with a distinguished label~\emph{id}, which is assumed to represent an
empty block without variables and statements.
As an illustration of the CFGs we consider, \cref{fig:cfg} depicts the
CFG of the workflow in \cref{ex:easychair}.

Let $(V,E)$ be the CFG of workflow $w$. We abuse notation, and we use
the proposition (i.e.~nullary predicate) $n$ to express whether the workflow is in node
$n\in V$. We denote by~$\cfgrels$ the set of these predicate symbols. 
The transition relation of the workflow is then expressed by the formula:
\[
    \fcfg(w) := \globally\, \big(\bigwedge_{n\in V} n \implies \nextstep\, (\bigvee_{n' \in \mathit{succ}(n)} n')\big)\\
\]
where $\mathit{succ}(n)$ is the set of the successors of the node $n$
in the CFG.
Furthermore, the workflow can never be in two states at once:
\[
    \fsanity(w) := \globally\, \big(\bigwedge_{n,n'\in V, n \neq n'} \neg (n \land n')\big)
\]

Workflow loops can but need not terminate, and thus the same holds for
workflow executions. A terminating behavior could be imposed
by requiring that the node $\lastnode$ is eventually reached, using
the formula $\eventually \lastnode$.
We do not impose this requirement.

\begin{figure}[t]
  \begin{center}
    \usetikzlibrary{automata}
\begin{tikzpicture}[->, node distance=1.7 cm]
  \node[state, initial, initial text={}] (0) {$n_0$};
  \node[state, right of=0] (1) {$n_1$};
  \node[state, right of=1] (2) {$n_2$};
  \node[state, right of=2] (3) {$n_3$};
  \node[state, above of=3] (4) {$n_4$};
  \node[state, right of=3] (5) {$\lastnode$};

  \path (0) edge node[below] {$b_1$} (1);
  \path (1) edge node[below] {$b_2$} (2);
  \path (2) edge node[below] {$b_3$} (3);
  \path (3) edge [bend right] node[right] {$b_4$} (4);
  \path (4) edge [bend right] node[left] {$b_5$} (3);
  \path (3) edge node[below] {$\mathit{id}$} (5);
  \path (5) edge [loop above] node {$\mathit{id}$} (5);
\end{tikzpicture}
  \end{center}
  \caption{CFG of the workflow in \cref{ex:easychair}.\label{fig:cfg}}
\end{figure}

\paragraph{\bf Initial state.}

Every workflow executes sequentially, thus starting at node $n_0$ of the workflows CFG, where $n_0$ is the entry point in the CFG.
There all relations in $\wfrels$ are empty.
Formally, this initial condition is expressed by the following formula:
\[ \finitial(w) := \big(n_0 \land \bigwedge_{n\in V\setminus\{n_0\}} \neg n\big) \land \bigwedge_{R\in\wfrels}\forall \bar{y}.\; \neg R(\bar{y}) \]

\paragraph{\bf Block execution.}

The basic step of our workflow language, which determines the
transition from the current time point to the next time point, is the
execution of one block.
As in~\cite{ATVA16}, we formalize this execution by characterizing
with an \fotl formula the interpretation of a predicate at the next
time point based on the interpretation of the relations at the current
time point.

For every block $b$ and every relation symbol~$R\in\wfrels$,
we construct a formula $\Phi_{b,R}(\bar{y})$
so that $R(\bar{y})$ holds after execution of block~$b$ iff
$\Phi_{b,R}(\bar y)$ holds before the execution of~$b$, with $|\bar{y}|=|\ar(R)|$.
The semantics of a block is then represented by the following formula:
\[
    \fblocks_w(b) := (n \land \nextstep n') \implies \bigwedge_{R \in \wfrels} \big(\forall\bar{y}. (\nextstep R(\bar{y})) \equ \Phi_{b,R}(\bar{y})\big)
\]
where $(n,n')\in E$ is the edge in the CFG with label~$b$.

For defining the formulas $\Phi_{b,R}$, we consider first a \wmay block~$b$, of the form $\wforall\;\bar x\ofType{\bar{s}}\ \wmay\ \wstmts$.
We assume, for simplicity, that each $R\in\wfrels$ is updated at
most once in a block (i.e. it occurs at most once in a statement of
block~$b$ on the right-hand side of~$\to$). 
For each $R\in\wfrels$, we let $\Phi_{b,R}(\bar y) :=$
\[
\left\{
\begin{array}{ll}
R(\bar y) & \text{if $R$ not updated}
\\
R(\bar y) \lor \exists \bar x\ofType{\bar{s}}.\; \theta \land \aux_i(\bar x) \land (\bar y = \bar u) 
& \text{if $\bar u$ added to $R$}
\\
R(\bar y) \land \neg \big(\exists \bar x\ofType{\bar{s}}.\; \theta \land \aux_i(\bar x) \land (\bar y = \bar u)\big) 
& \text{if $\bar u$ deleted from $R$}
\end{array}
\right.
\]
where block $b$'s statement that updates $R$ has the
form~$\theta\to R \wpm \bar u$, $i$ is the index of the block $b$ in
the linearisation of the workflow, and it is assumed that
$\bar{x}\cap\bar{y}=\emptyset$.
The definition of $\Phi_{b,R}$ when $b$ is a non-\wmay block is
similar, except that the $\aux_i(\bar{x})$ conjuncts are omitted.

If the same relation is modified multiple times in a block, the updates take place sequentially.
This is expressed by building the formulas $\Phi_{b,R}$ inductively, similarly as above: $\Phi_{b,R}$ up to statement~$j$ is obtained by replacing $R(\bar y)$ with $\Phi_{b,R}$ up to statement~$j-1$. We omit the precise formalization.

We denote by $\lirels$ the $\aux_i$ predicate symbols used in the
$\Phi_{b,R}$ formulas. These symbols denote relations that contain
\emph{low input} (i.e.~non-confidential input) to the workflow.
For instance, in \cref{ex:easychair} we have
$\lirels=\{\aux_1,\aux_2,\aux_5\}$.

The semantics of the workflow execution is then captured by the following formula:
\[
    \fblocks(w) := \globally\, \bigwedge_{b\in \blocks(w)}\fblocks_w(b).
\]
where $\blocks(w)$ is the set of $w$'s blocks.

\begin{example} The execution semantics of the block $(b_2)$ of the workflow from Example~\ref{ex:easychair} is given by the following formula:
\begin{align*}
& (n_1 \land \nextstep n_2) \implies  
\\
&  \quad \big(\forall y_1,y_2.\; \nextstep \Assign(y_1,y_2) \equ 
  \Assign(y_1,y_2)\; \lor
\\
&  \quad\qquad (\exists x,p.\ \aux_1(x,p) \land \neg\Conf(x,p) \land y_1=x \land y_2=p)\big)\; \land \\
& \quad \big(\forall y_1,y_2.\; \nextstep \Conf(y_1,y_2) \equ \Conf(y_1,y_2)\big) \land \ldots 
\end{align*}
The first conjunct of the above consequent can be rewritten into the logically equivalent formula
\[\forall x,p.\ \nextstep \Assign(x,p) \equ \Assign(x,p) \lor \aux_1(x,p) \land \neg\Conf(x,p)\] 
by substituting $x$ and $p$ by $y_1$ and $y_2$ respectively, and then
renaming $y_1$ and $y_2$ back to $x$ and $p$. We note that this
formula matches well the syntax of the block~$(b_2)$.
The mentioned simplification cannot be performed in general,
but only for a class of workflows, see Section~\ref{sec:wellformed}.
\end{example}

We note that a $\wforall\ \bar{x}\ \wmay$ block with statements of the
form $\theta_i \to R_i \wpm \bar u_i$, can be seen as an abbreviation
of a non-\wmay block with statements of the form
$\aux(\bar{x}) \land \theta_i \to R_i \wpm \bar u_i$, for some
predicate symbol~$\aux\notin\wfrels$.
Note also that for an atom~$\Oracle(\bar{t})$ occurring in some
guard~$\theta_i$, the arity of $\Oracle$ need not be~$|\bar{x}|$.
We use the abbreviated \wmay form to emphasize the subtle differences
between the two kinds of non-workflow relations.

\paragraph{\bf Summary.}
The complete specification $\fwf(w)$ of the workflow is a conjunction of the several parts described previously --- the control flow graph, the initial state, and the semantics of the transitions between time points.
\[
\fwf(w) := \fcfg(w) \land \fsanity(w) \land \finitial(w) \land \fblocks(w)
\]
Note that the formula~$\fwf(w)$ is expressed over the signature $\sig'$ obtained from $\sig$ by extending it with relation symbols $n\in\cfgrels$ and $\aux_i\in\lirels$.

For any given workflow $w$ over a signature $\sig$, its semantics
$\sem{w}$ consists of all temporal structures $\tstruct$ over $\sig'$
that satisfy $\fwf(w)$.
A workflow~$w$ \emph{satisfies} a closed \fotl formula $\phi$, denoted
$w \models \phi$, iff $\tstruct,\nu \models \phi$ for any
$\tstruct\in\sem{w}$, and any valuation $\nu$.
We have:

\begin{theorem}\label{thm:workflows_enc}
  Given a workflow $w$, a \fotl formula $\phi_w$ can be built in
  polynomial time so that for every \fotl formula $\phi$, it holds
  that $w\models\phi$ iff $\phi_w\implies\phi$ is valid.
\end{theorem}
In fact, as such $\phi_w$ we may choose the formula $\fwf(w)$.

\subsection{Non-omitting Workflows}\label{sec:wellformed}

We call a workflow \emph{non-omitting} iff for each of its blocks 
\[
\begin{array}{l}
\wforall\;\bar x\ofType{\bar s}\ [\wmay]     \\
\quad
\begin{array}[t]{l}
\theta_1\to R_1 \wpm \bar u_1; \\
\ldots  \\
\theta_n\to R_n \wpm \bar u_n \\
\end{array}
\end{array}
\]
we have $\fv(\bar{u}_i) = \bar x$ and $\theta_i$ is quantifier-free, for each~$i\in\{1,\dots,n\}$.

For a non-omitting workflow~$w$, we can replace all existentially
quantified variables inside the $\Phi_{b,R}$ formulas by their
respective values, and remove the existential quantifiers.  Thus, for
any block~$b$, as all guards of~$b$ are quantifier-free, $\Phi_{b,R}$
becomes quantifier-free, for all $R\in\wfrels$.
It follows that the formula $\fwf(w)$ can be brought into the \bsfotl
fragment. 
Note that the thus simplified $\fwf(w)$ formula contains no
existential quantifiers. Furthermore, all its universal quantifiers
are either not under a temporal operator (in the case of the
$\finitial(w)$ subformula) or under the $\globally$ temporal operator
(in the case of the $\fblocks(w)$ subformula). Therefore the
simplified $\fwf(w)$ formula can be put in prenex normal form having a
quantifier prefix consisting of only universal quantifiers. As a side
remark, this means that $\neg\fwf(w)$ can also be brought into
\bsfotl.

\begin{theorem}\label{thm:workflows_fo_sat}
  It is decidable for a non-omitting workflow~$w$ and a formula $\phi$
  in \bsfotl whether or not $w\models\neg\phi$ holds.
\end{theorem}

This means that if the set of all \emph{bad} behaviors can be
expressed by a formula $\phi$ in \bsfotl, then absence of bad
behaviors can be checked for non-omitting workflows.
The theorem follows from Theorems~\ref{thm:workflows_enc}
and~\ref{thm:decidability}. Indeed, it is sufficient to check whether
$\fwf(w)\land\phi$ is unsatisfiable. This can be done, since both
conjuncts can be brought into \bsfotl, and thus the conjunction itself
too.

The following theorem shows that the decidability result from Theorem~\ref{thm:workflows_fo_sat}
cannot be lifted to arbitrary workflows.
\begin{theorem}\label{thm:omitting_undecidable}
  It is undecidable for a workflow $w$ and a formula $\phi$
  in \bsfotl whether or not $w\models\neg\phi$ holds.
\end{theorem}

\begin{proof}
  We prove the theorem by reducing the \emph{periodic tiling problem} to our workflow setting. 
  The tiling problem was first mentioned in~\cite{wang1990dominoes} and has first been shown undecidable by \citeauthor{berger1966undecidability} in~\cite{berger1966undecidability}. Its closely related variant, the periodic tiling problems has also been proven undecidable by multiple authors --- for an overview see~\cite{jeandel2010periodic}.
  We now briefly recall the definition of the problem.

  Given a set of $k$ tile types $T = \{ T_i \mid 0 \leq i < k \}$ as well as horizontal and vertical compatibility relations $\xcomp\subseteq T \times T$ and $\ycomp \subseteq T \times T$, 
  a \emph{tiling} is a function $f(x,y):\Nat \times \Nat \rightarrow T$ such that whenever two tiles are adjacent, they have to respect the compatibility relations:
  \begin{equation*}
  \begin{aligned}
    \forall x,y.\ \xcomp(f(x,y), f(x+1,y)),\\
    \forall x,y.\ \ycomp(f(x,y), f(x,y+1)).
  \end{aligned}
  \end{equation*}
  A tiling is \emph{periodic} if there exist horizontal and vertical periods $p_x$ and $p_y$, such that
  \begin{equation*}
  \begin{aligned}
    \forall x,y.\ f(x,y) = f(x+p_x,y),\\
    \forall x,y.\ f(x,y) = f(x,y+p_y).
  \end{aligned}
  \end{equation*}
  The periodic tiling problem is to find out for a given set of tile types and its compatibility relations, if there exists a periodic tiling.\footnote{The original formulation used just a single period $p$ in both directions. 
  We use independent periods to have less complicated constructions. 
  We note that given a periodic tiling $t$ with periods $p_x$ and $p_y$ it is easy to construct a periodic tiling $t'$ with $p_x' = p_y' = (p_x * p_y)$.
  The original problem also did not consider compatibility relations, but edges of the same color. Again this makes our constructions easier and is easily transformed into a solution of the original setting.}

  We will now proceed to show how to encode this problem in our workflow setting.
  We note that to find a periodic tiling, it is enough to find the periods $p_x$, $p_y$, and the values $f(x,y)$ for $0 \leq x < p_x$ and $0 \leq y < p_y$, such that they are also compatible at borders:
  \begin{equation*}
  \begin{aligned}
    \forall y.\ \xcomp(f(p_x,y), f(0,y)),\\
    \forall x.\ \ycomp(f(x,p_y), f(x,0)).
  \end{aligned}
  \end{equation*}
  We thus see a periodic tiling as a table with rows referring to
  points on the $y$-axis and columns referring to points on the
  $x$-axis.

  We build next a workflow $w$ and a formula $\phi$ such that $w\models\phi$ iff there is a periodic tiling for $(T,\xcomp,\ycomp)$. We use the following signature:
  \[
    \sig = \big( \{A\}, \{ \afirst, \alast \}, \{ Q,\adjy,\reachy,T'_0,\ldots,T'_{k-1} \}, \ar \big)
  \]
  Intuitively, time points refer to the rows of the tiling, while agents refer to its columns. 
  We explain next the role of the constant and relation symbols. The $k$
  unary relations $T'_i$, with $0 \leq i < k$, are used to encode the
  tiling function as follows: if $T'_i(a_j)$ holds at time point~$t$,
  for some particular agent~$a_j$, then the tiling function is
  $f(t,j) = T_i$. How the agent~$a_j$ is determined is explained
  later.
  There are two constant agents $\afirst$ and $\alast$ which are used to name the first and last row of the tiling.
  The nullary relation~$Q$ encodes the last column of the tiling.
  %
  The predicate~$\adjy(a,a')$ expresses that the row named by~$a'$ is directly below the row named by~$a$.
  %
  Only $\reachy$ is a workflow relation; thus, initially (i.e.~at time point~0) it is empty.
  There is a single sort, the agent sort~$A$. 

  We let $w$ be the following workflow. It is used to compute all reachable parts of the adjacency relation \adjy starting from the initial agent $\afirst$:

  \begin{align*}
    &\wforall .\ \ttrue \to \reachy \wadd (\afirst)\\
    &\wloop\\
    &\qquad \wforall\ a,a'.\ \reachy(a) \land \adjy(a,a') \to \reachy \wadd(a')
  \end{align*}

  To encode the rest of the tiling requirements, we use a conjunction of \bsfotl formulas, where $i,j$ implicitly range over the elements in $\{0,\dots,k-1\}$:

\noindent  All agents always have exactly one tile assigned at each point in time (\cref{eq:omitting_assign1,eq:omitting_assign2}).
      \begin{equation}\label{eq:omitting_assign1}
        \LTLglobally \forall a.\ \bigvee_i T'_i(a)
      \end{equation}
      \begin{equation}\label{eq:omitting_assign2}
          \LTLglobally \forall a.\ \bigwedge_{i \neq j} T'_i(a) \implies \neg T'_j(a)
      \end{equation}  

\noindent A time point will be reached state where $Q$ holds (\cref{eq:omitting_Q1}) and it will only hold once (\cref{eq:omitting_Q2}).
      \begin{equation}\label{eq:omitting_Q1}
          \LTLnext \LTLfinally Q
      \end{equation}
      \begin{equation}\label{eq:omitting_Q2}
          \LTLglobally\ (Q \implies \LTLnext \LTLglobally \neg Q)
      \end{equation}

\noindent Two adjacent time points need to be assigned $x$-compatible tiles (\cref{eq:omitting_lrcomp}).
    Also, the right border of the tiling should be $x$-compatible to the left, i.e. the time point where $Q$ holds should be compatible to the starting time point (\cref{eq:omitting_lrperiod}).
      \begin{equation}\label{eq:omitting_lrcomp}
        \forall a.\ \LTLglobally\ (\bigvee_{i,j:\, \xcomp(T_i,T_j)} T'_i(a) \land \LTLnext T'_j(a) )
      \end{equation}

      \begin{equation}\label{eq:omitting_lrperiod}
        \forall a.\ \bigvee_{i,j:\, \xcomp(T_i,T_j)} T'_j(a) \land \LTLfinally \big( Q \land T'_i(a) \big)
      \end{equation}

\noindent Two adjacent agents need to be assigned $y$-compatible tiles (\cref{eq:omitting_tbcomp}).
    The last agent should be reachable from the first via $\adjy$ relations.
    We cannot express this fact in pure \bsfotl, so we will use the relation \reachy computed by the workflow (\cref{eq:omitting_tbreach}).
    The last agent should also be $y$-compatible to the first (\cref{eq:omitting_tbperiod}).
      \begin{equation}\label{eq:omitting_tbcomp}
        \LTLglobally \forall a,a'.\ (\LTLfinally \adjy(a,a')) \implies \bigvee_{i,j:\,\ycomp(T_i,T_j)} T'_i(a) \land T'_j(a')
      \end{equation}

      \begin{equation}\label{eq:omitting_tbreach}
        \LTLfinally \reachy(\alast)
      \end{equation}

      \begin{equation}\label{eq:omitting_tbperiod}
        \LTLglobally\ (\bigvee_{i,j:\,\ycomp(T_i,T_j)} T'_i(\alast) \land T'_j(\afirst))
      \end{equation}

      Let $\phi$ be the conjunction of
      \cref{eq:omitting_assign1,eq:omitting_assign2,eq:omitting_Q1,eq:omitting_Q2,eq:omitting_lrcomp,eq:omitting_lrperiod,eq:omitting_tbperiod,eq:omitting_tbreach,eq:omitting_tbcomp}. Note that $\phi$ can be brought in \bsfotl.
      We show next that $w\models\phi$ iff there is a periodic tiling for $(T,\xcomp,\ycomp)$.

    Let $\tstruct\in\sem{w}$ such that $\tstruct\models \phi$. We construct a tiling as follows. 
    As $\tstruct$ satisfies the formula~(\ref{eq:omitting_tbreach})
    and the formulas encoding the first and second blocks of the
    workflow, it follows that there is a sequence $(t_0,\dots,t_n)$ of
    time points with $n>0$ and $t_0=0$, and a sequence
    $(a_0,\dots,a_n)$ of elements of the universe such that
    $a_0=\afirst$, $a_n=\alast$, $\reachy(a_i)$ holds at time
    point~$t_i$, for all $i$ with $0\leq i\leq n$, and
    $\adjy(a_i,a_{i+1})$ holds at time point $t_i$, for all $i$ with
    $0\leq i< n$.
    Then, we set $p_x$ to the time point where $Q$ holds and $p_y$ to $n$.
    For $0 \leq t < p_x$ and $0 \leq j < p_y$, let $f(t,j)$ be $T_i$ iff $T'_i(a_j)$ holds at time point~$t$.
    It is easy to see that $f$ satisfies the compatibility relations and that any given tiling can be transformed into a model of~$\phi$.
\end{proof}

\section{Hyperproperties}
\label{sec:prop}

In this section, we show how to formalize and verify security
properties of workflows. We focus on non-interference
properties~\cite{GoguenMeseguer82}, which are
hyperproperties~\cite{ClarksonS10}. To specify such properties we use
the first-order extension of HyperLTL~\cite{clarkson2014temporal}
presented in~\cite{ATVA16}.
HyperLTL can relate multiple traces and it is thus well suited to
express not only trace properties, but also hyperproperties.
The first-order extension is needed in the presence of an unbounded
number of agents. Furthermore it allows for more fine-grained
policies.

\subsection{HyperFOLTL}
\label{sec:hyperltl}

For presenting the syntax and semantics of the logic, we follow~\cite{ATVA16}.

\paragraph{\bf Syntax}

Let $\sig=(\sorts,\csts,\rels,\ar)$ be a signature, and let $\Pi$ be a set of
\emph{trace variables} disjoint from the set $\vars$ of first-order
variables.
Let $\rels_{\Pi} = \{ R_\pi \mid R \in \rels, \pi \in \Pi\}$ and
$\sig' = (\sorts,\csts,\rels_\Pi,\ar')$ be the signature with
$\ar'(R_\pi)=\ar(R)$, for any $R\in\rels$ and $\pi\in\Pi$.

\hfotl extends \fotl as follows.
\hfotl \emph{formulas} over~$\sig$ and~$\Pi$ are then generated by the
following grammar:
\[
\psi ::= \exists\pi.\ \psi \mid \neg \psi \mid \phi\\
\]
where $\pi\in\Pi$ is a trace variable and $\phi$ is a \fotl formula over~$\sig'$.
Universal trace quantification is defined as $\forall\pi.\psi :=
\neg\exists\pi.\neg\psi$.
\hfotl formulas thus start with a prefix of trace
quantifiers consisting of at least one quantifier and then continue
with a subformula that contains only first-order quantifiers, no trace
quantifiers.  
As for \fotl, a formula without free first-order and
trace variables is called \emph{closed}.

\paragraph{\bf Semantics}

The {semantics} of a \hfotl formula $\psi$ is given with respect to
a set $\mathcal{T}$ of temporal structures, a valuation $\alpha: \mathcal
V \rightarrow U$ of the first-order variables, and a valuation
$\beta: \Pi \rightarrow\mathcal{T}$ of the trace variables.
The \emph{satisfaction} of a \hfotl formula $\psi$, denoted by
$\mathcal{T}, \alpha, \beta \models \psi$, is then defined as follows:
\[
\begin{array}{l@{\hspace{1em}}c@{\hspace{1em}}l}
{\mathcal{T}}, \alpha,\beta \models \exists \pi.\  \psi & \text{iff} & \text{${\mathcal{T}}, \alpha,\beta[\pi \mapsto t] \models \psi$, for some $t \in {\mathcal{T}}$,}\\
{\mathcal{T}}, \alpha,\beta \models \neg \psi & \text{iff} & {\mathcal{T}}, \alpha,\beta \not\models \psi, \\
{\mathcal{T}}, \alpha,\beta \models \phi  & \text{iff} &  \tstruct, \alpha \models \phi,\\
\end{array}
\]
where $\psi$ is an \hfotl formula, $\phi$ is an \fotl
formula, and the temporal structure~$\tstruct$ is such that for all
$R\in\rels$, $i\in\Nat$, and, $\pi\in\Pi$, the interpretation
$R_\pi^{\struct_i}$ is $R^{\beta(\pi)(i)}$ if $\pi$ in the domain of
$\beta$, and $\emptyset$ otherwise.

A \hfotl formula $\psi$ is \emph{satisfiable} iff there exists a
set~$\mathcal{T}$ of temporal structures and valuations $\alpha$ and
$\beta$ s.t.~$\mathcal{T},\alpha,\beta \models \psi$.

\begin{example}\label[example]{ex:obsdet}
  Observational determinism~\cite{ObservationalDeterminism} of programs
  can be formalized by the following \hfotl formula
  \[
    \forall \pi,\pi'.\, (\globally \forall x.\, I_\pi(x) \equ I_{\pi'}(x)) 
    \implies (\globally \forall y.\, O_\pi(y) \equ O_{\pi'}(y)),
  \]
  where $I(x)$ denotes that $x$ is a low input to the program, while
  $O(y)$ denotes that $y$ is a low output. The inputs and outputs are
  classified as low or high with respect to the clearance level of
  some particular user. The formula states that, on any two program
  executions, if the low inputs are always the same, then the low
  outputs are also always the same. That is, from a low user point of
  view, the observable behavior of the program is only determined by
  its inputs.
\end{example}

We will adapt this non-interference notion to the workflow setting in
\cref{sec:noninterference}.
We refer to~\cite{clarkson2014temporal} for the formalization in
HyperLTL of other hyperproperties.

\paragraph{\bf Decidability}

We will consider the fragment of \hfotl, named
$\exists^*_\pi\forall^*_\pi\bsfotl$, that consists of all formulas of
the form
$\exists \pi_1,\dots\pi_k.\, \forall\pi'_{1}\dots\pi'_{\ell}.\, \phi$ with
$k\geq 0$, $\ell\geq 0$, and $\phi$ an \fotl formula in \bsfotl.

We first remark that by seeing trace variables as first-order variables
of a new sort~$T$ --- the trace sort, \hfotl formulas can be
faithfully encoded by \fotl formulas. By this we mean that, for any
closed \hfotl formula $\psi$, there is a closed \fotl formula~$\phi$
such that we can translate models of $\psi$ into models of $\phi$ and
vice-versa.
The formula $\phi$ is obtained by replacing trace quantification
$Q\pi$ to first-order quantification $Q\pi\ofType{T}$, for
$Q\in\{\exists,\forall\}$, and predicates $R_\pi(\bar{u})$ with
predicates $R'(\pi,\bar{u})$. Note that $\psi$ and $\phi$ are
formulas over slightly different signatures.
The translation between models is straightforward. For instance, if
$\tstruct$ is a temporal structure that satisfies~$\phi$, then the
corresponding set~$\mathcal{T}$ of temporal structures that
satisfies~$\psi$ consists of temporal structures obtained by
projecting a predicate's interpretation on the predicate's non-trace
arguments, for each of the values of the trace universe~$U_T$
of~$\tstruct$, i.e.  $\mathcal{T} = \{\tstruct_t \mid t\in U_T\}$ and
$R^{\struct_{t,i}} = \{\bar{a} \mid (t,\bar{a})\in R'^{\tstruct_i}\}$,
for each $R\in\rels$, $t\in U_T$, and~$i\in\Nat$.

As a consequence of the previous discussion, and as a corollary of
Theorem~\ref{thm:decidability}, we obtain the following results.

\begin{theorem}\label{thm:hltl_bs_fragment}
  The following statements hold.
  \begin{enumerate}
  \item Every \hfotl formula can be translated into an
    equi-satisfiable \fotl formula.
  \item Satisfiability of formulas in $\exists^*_\pi\forall^*_\pi\bsfotl$ is decidable.
  \end{enumerate}
\end{theorem}

\paragraph{\bf Workflow satisfaction}

A workflow~$w$ \emph{satisfies} a closed formula \hfotl $\psi$, denoted
$w \models \psi$, iff $\sem{w}, \alpha,\beta \models \psi$ for the
empty assignments $\alpha$ and~$\beta$.

\begin{theorem}\label{thm:workflows_hltl_sat}
  Let $w$ be a workflow and $\psi$ a \hfotl formula.
  Then the following statements hold.
  \begin{enumerate}
  \item An $\fotl$ formula $\psi'$ can be constructed in polynomial
    time so that $w\models\neg\psi$ iff $\psi'$ is unsatisfiable.
  \item If $w$ is non-omitting and $\psi$ is in
    $\exists^*_\pi\forall^*_\pi\bsfotl$, then it is decidable whether
    or not $w\models\neg\psi$ holds.
  \end{enumerate}
\end{theorem}

\begin{proof}
  Assume $\psi$ has the form
  $Q_1\pi_1\dots Q_k\pi_k.\,\phi$,
  where the trace quantifiers are partitioned into a set $E$ of existential quantifiers and a set $A$ of universal quantifiers.
  Then $w\models\neg\psi$ is equivalent with the validity of the following \hfotl formula
  $$\bar{Q}_1\pi_1\dots \bar{Q}_k\pi_k.\,
  \Big(\bigwedge_{\bar{Q}_i \in E}\fwf(w)_{\pi_i}\Big)\land \Big( \Big( \bigwedge_{\bar{Q}_i \in A}\fwf(w)_{\pi_i}\Big)\rightarrow \neg\phi\Big),$$ 
  where $\fwf(w)_\pi$ is $\fwf(w)$ with each predicate symbol $R$
  replaced by the predicate symbol~$R_\pi$, and $\bar{Q}$ is $\exists$ if $Q$ is $\forall$ and vice-versa.
  The formula~$\psi'$ is then the \fotl encoding of the following \hfotl formula
  $$Q_1\pi_1\dots Q_k\pi_k.\,
  \Big(\bigwedge_{Q_i \in A}\fwf(w)_{\pi_i}\Big) \to \Big( \Big( \bigwedge_{Q_i \in E}\fwf(w)_{\pi_i}\Big)\land \phi\Big).$$ 

  From Theorem~\ref{thm:hltl_bs_fragment}, to prove the second
  statement, it is sufficient to show that the previous \hfotl
  formula, which we call $\psi_1$, can be brought in the
  $\exists^*_\pi\forall^*_\pi\bsfotl$.
  By assumption, we have that the trace quantifier prefix of $\psi_1$
  is of the form $\exists^*_\pi\forall^*_\pi$ and that $\phi$ is in
  the $\bsfotl$ fragment. Also, since $w$ is non-omitting, then both
  $\fwf(w)$ and $\neg\fwf(w)$ can be brought in the $\bsfotl$
  fragment, as remarked in \cref{sec:wellformed}. Thus all conjuncts
  in the following \fotl formula can be brought in the $\bsfotl$
  fragment
  \[
  \Big(\bigwedge_{Q_i \in A}\neg\fwf(w)_{\pi_i}\Big) \lor \Big( \Big( \bigwedge_{Q_i \in E}\fwf(w)_{\pi_i}\Big)\land \phi\Big)
  \] 
  This means that the formula itself can be put into $\bsfotl$ and
  thus $\psi_1$ can be brought into
  $\exists^*_\pi\forall^*_\pi\bsfotl$.
\end{proof}

\subsection{Non-interference in workflows}\label{sec:noninterference}

As we have defined it, the workflow keeps track of the state of all relations of all agents.
However, security policies are meant to allow access to classified information to just some of the users of the system while denying it to others.
For this, we need to specify how an agent interacts with the workflow and reason about his knowledge and possible interactions with the system.

In the running example, members of the PC can use a conference management system 
to specify conflicts, read the reviews that other members have provided, provide their own reviews, etc.
As an example property, we will formalize that no member of the PC gains any information about papers that he declared a conflict of interest with. 

Following~\cite{ATVA16}, we present a variant of \emph{non-interference}
suitable for these properties on workflows
by adapting the notion of observational determinism from \cref{ex:obsdet} to explicitly take
into account the knowledge and behavior of participating agents.

Non-interference in general is a strong specification of the valid information flows in a system.
It uses a classification of all inputs and outputs to a system into ``high'' security and ``low'' security inputs~\cite{GoguenMeseguer82}.
In our setting, these notions of input and output are specific to an agent and his interactions with the workflow.
input to model $a$'s interactions with the workflow.
We call a workflow \emph{non-interferent}, iff for any agent~$a$, his observations do not 
depend on the inputs which are ``high'' for~$a$ in any way.

\paragraph{\bf Agent Model}

It has been observed in \cite{ATVA16}, that 
non-interference in workflows can only reasonably be argued about, if 
meaningful assumptions on the behavior of agents are provided.

In order to specify such assumptions, we make the convention that in any relation recording
an agent's knowledge or interaction, this agent appears in the first
argument of the relation.
Formally, we classify all sorts into agent sorts and data sorts.
Moreover, we require that the arity $(s_1,s_2,\dots)$ of every relation
$R\in\wfrels\cup\lirels\cup\hirels$ is non-nullary and is such that
$s_1$ is an agent sort. This restriction, while not strictly necessary, allows us to present the results in this section in a much cleaner way.

An agent provides observable input to the workflow system by
choosing to execute (or to not execute) \wmay-blocks for specific data.
Such input is low input, formalized through the predicates $\aux\in\lirels$.
The property that at a given time point, all low inputs for a given agent~$a$ are equal on traces $\pi, \pi'$ is formalized as:
\begin{align*}
& \lowinputs_{\pi, \pi'}(a) := & \\
& \quad
  \bigwedge_{\aux\in\lirels} \big(\forall \bar x.\, \aux_{\pi}(a,\bar x) \equ \aux_{\pi'}(a,\bar x)\big),
\end{align*}
where, for each $\aux$ predicate, the sequence $\bar{x}$ has the same
length as its arity minus 1.
Input provided by the environment is considered high input. It is
formalized through predicates $\Oracle\in\hirels$.

An agent can observe all tuples in which it is mentioned in the first argument. 
The property that, at a time point, all observations of a given
agent~$a$ are the same on two given traces~$\pi$ and~$\pi'$ is
formalized by the following formula:
\[
\lowoutputs_{\pi, \pi'}(a) := \bigwedge_{R \in \wfrels}(\forall \bar x.\, R_\pi(a,\bar x) \equ R_{\pi'}(a,\bar x) )
\]

\paragraph{\bf Agent Behavior}
The behavior of the workflow as seen by one of the agents, depends on the actions of all other agents.
If agents have the power to behave arbitrarily, there will be spurious counterexample traces
to confidentiality where an agent chooses to let his actions depend on confidential data --- 
which he could not even access.
Here, we consider two meaningful agent models which restrict the behavior of agents 
across different executions.

The simpler agent model considers \emph{\textstubborn} agents.
An agent is called \textstubborn if, even when told information that is confidential to another agent, 
he will not choose his actions depending on this information.
Thus, anyone observing the behavior of a \textstubborn agent will not be able to conclude 
anything about confidential data.
Technically, this amounts to saying that his choices are independent of the chosen trace.
For a pair of traces $\pi,\pi'$, the behavior of a \textstubborn agent is therefore specified in \hfotl by the following formula:
\[
 \stubborn_{\pi,\pi'}(a) := \globally \lowinputs_{\pi,\pi'}(a)
\]

A more intricate model of agent behavior considers \emph{\textcausal} agents.
An agent is called \textcausal if his actions may depend on his observations.
As a result, a \textcausal agent can subtly change his behavior depending on the data 
that he gained access to.
As an example, a \textcausal agent could indicate the acceptance of a paper to someone else
either by explicitly telling it to someone or by commenting to another paper or refraining from it.
For a pair of traces~$\pi,\pi'$, this behavior is specified in \hfotl by the following formula.
\begin{align*}
& \causal_{\pi,\pi'}(a) := \\
& \qquad \lowinputs_{\pi,\pi'}(a) \wuntil \neg \lowoutputs_{\pi, \pi'}(a)
\end{align*}
We remark that the causal agent model subsumes the \textstubborn agent model and is less constraining on the behavior of the individual agents, which leads to more intricate information flow violations.

We remark that the formula $\forall a. \causal(\pi,\pi',a)$ is not expressible in \bsfotl, as it has an $\forall\exists$ quantifier structure.
In case, however, that we consider a fixed upper bound on the number of \textcausal agents, 
the corresponding formula is in the \bsfotl fragment.
For instance for at most two agents, we can use the following formula:
\[
  \exists a_1, a_2.\, \causal_{\pi,\pi'}(a_1) \land \causal_{\pi,\pi'}(a_2)
\]
Considering an upper bound on the number of \textcausal agents is a realistic setting, 
as it allows to verify the system for attacks by coalitions up to a given size.

\paragraph{\bf Declassification}

In general, all external input data to the workflow, i.e. all relations in $\hirels$ are 
considered as high input.
However, it often needs be possible that an agent can learn something about  
the high input data, depending on the scenario.
This is also apparent in the conference management example. There, 
it is necessary for a reviewer to be able to read at least some reviews, namely,
the reviews for papers he himself is assigned to --- although reading these might be
illegitimate for others.

To model declassification, we assume a formula $\phi_{\Oracle}$ for each relation 
$\Oracle$ in $\hirels$. This formula encodes a declassification condition that describes which $\Oracle$ tuples represent declassified information, for any given agent.
Initial high inputs for $a$ therefore should only be equal on traces $\pi, \pi'$ if they are 
declassified for agent $a$. 
Technically, this property is formalized by:
\begin{align*}
\highinputs_{\pi, \pi'}(a) := \hspace*{8em}\\
\LTLglobally \bigwedge_{\Oracle\in\hirels} \forall \bar y.\, \left(
  \begin{aligned}
    & (\phi_{\Oracle,\pi}(a,\bar y) \lor \phi_{\Oracle,\pi'}(a,\bar y)) \\
    & \implies (\Oracle_\pi(\bar y) \equ \Oracle_{\pi'}(\bar y)) 
  \end{aligned}
  \right)
\end{align*}
By the notation $\phi_{\Oracle}(a,\bar y)$ we mean that the free
variables of the formula~$\phi_{\Oracle}$ are among the variables
$a$ and those in~$\bar{y}$.
For our running example, we use $\phi_\Oracle(a, x, p, r) := \neg \Conf(a,p)$.

\paragraph{\bf Control Flow}
The structure of the control flow graph and the current position of the workflow
(i.e. the state of all relations in~$\cfgrels$) are considered as \emph{low} input. 
This serves the intuition that the non-determinism in the workflow is resolved by some external control.
For instance, the PC chair of the conference management system may terminate the submission loop.
This assumption is formalized by the following formula:
\[
\samecfg_{\pi,\pi'} := \LTLglobally \bigwedge_{n\in\cfgrels} n_{\pi} \equ n_{\pi'}
\]

\paragraph{\bf Putting it all together}

Assume that there are at most $k\geq 0$ \textcausal agents with all other agents being \textstubborn.
\emph{Non-interference with Declassification} is then expressed in \hfotl by the 
following formula: 
\begin{align*}
  & \forall \pi,\pi'.\,
  \Big(\exists a_1,\dots,a_k.\,
       \big( \bigwedge_{i=1}^k \causal_{\pi,\pi'}(a_i) \big) \ \land 
  \\ & \hspace*{10em}
       \big( \forall a.\, (\bigwedge_{i=1}^k a \neq a_i) \to \stubborn_{\pi,\pi'}(a) \big)\Big)
  \\ & \hspace*{3em} \ \land \samecfg_{\pi,\pi'}
  \\ & \hspace*{3em}
       \implies \forall a.\, \noninter_{\pi,\pi'}(a)
\end{align*}
where $\noninter_{\pi,\pi'}(a) :=$
\begin{align*}
 & \ \big(
 \big(\LTLglobally \lowinputs_{\pi,\pi'}(a)\big)\ \land\\
& \phantom{\ \big(} \highinputs_{\pi,\pi'}(a) \big)\\
& \implies \LTLglobally \lowoutputs_{\pi,\pi'}(a).
\end{align*}

\begin{table}
  \caption{A counterexample to non-interference.}
  \label{tab:counterexample}
  \begin{tabular}{|c|l|l|l|}
    \hline
    block & relation & $\pi$ & $\pi'$ 
    \\
    \hline\hline
    $(b_1)$ & $\Conf$ & \multicolumn{2}{|c|}{$(a_1,p_1)$}
    \\
    \hline
    $(b_2)$ & $\Assign$ & \multicolumn{2}{|c|}{$(a_1,p_2), (a_2,p_2), (a_2,p_1)$}
    \\
    \hline
    \multirow{2}{*}{$(b_3)$} & \multirow{2}{*}{$\Review$} & $(a_2,p_1,r_{21})$ & 
    \\
    & & $(a_2,p_2,r_{22})$ & $(a_2,p_2,r_{22})$
    \\
    \hline
    \multirow{3}{*}{$(b_4)$} & \multirow{3}{*}{$\Read$} &
    $(a_1,a_2,p_2,r_{22})$ & $(a_1,a_2,p_2,r_{22})$
    \\ & & $(a_2,a_2,p_2,r_{22})$ & $(a_2,a_2,p_2,r_{22})$
    \\ & & $(a_2,a_2,p_1,r_{21})$ & 
    \\
    \hline
    $(b_5)$ & $\Review$ & $(a_2,p_2,r_{21})$ & 
    \\
    \hline
    \multirow{2}{*}{$(b_4)$} & \multirow{2}{*}{$\Read$} & $(a_1,a_2,p_2,r_{21})$ & 
    \\ & & $(a_2,a_2,p_2,r_{21})$ & 
    \\
    \hline
  \end{tabular}
\end{table}

\begin{example}\label{ex:counterex}
Coming back to the workflow in \cref{ex:easychair}, we check if the non-interference property holds.

When all agents are \textstubborn, we find that non-interference is satisfied for the given workflow.
This result indicates that there is no way for any agent to learn confidential information without having a conspirator helping him.

The result is different when there is at least one \textcausal agent. In this case we find the following counterexample:
Assume two PC members $a_1$ and $a_2$ where $a_1$ is \textstubborn and $a_2$ is \textcausal. The non-interference property is stated for~$a_1$.
There are two papers $p_1$ and $p_2$.
First, $a_1$ declares a conflict with $p_1$, so in the rest of the workflow he 
should not be able to observe a difference between two executions of the workflow, regardless of which reviews $p_1$ receives.
Both agents get assigned to $p_2$. In addition, $a_2$ gets assigned to $p_1$ and writes a review for it.
At this point, $a_2$ can observe at least one review for $p_1$, so he can deviate his behavior on the two executions.
The next step is the discussion phase.
In the first step, $a_2$ reads all reviews of $p_1$.
In the next step, $a_2$ adjusts his reviews of $p_2$ to mirror the reviews of $p_1$.
Then, in the next iteration, $a_1$ will read the differing reviews of $p_2$ 
and learn about the result of $p_1$, the paper he initially declared a conflict with.

\cref{tab:counterexample} formalizes the counterexample. It shows the
tuples that are added to the updated relation after the execution of
each block. Note that the workflow updates only one relation per
block and there are no removals.
The reviews for $p_2$ cannot differ (in the two traces) directly after the execution of the block~$(b_3)$ since the declassification condition states that tuples in $\Oracle$ can only differ when they are of the form $(x,p_1,r)$. 
However, as $a_2$ can observe his own reviews for $p_1$, his choices can start to differ after~$(b_3)$ is executed; concretely, they will differ when block~$(b_5)$ is executed.
In the last two rows, any value for $r$ (except $r_{22}$) would result in a
counter-example; we use $r_{21}$ to suggest that $a_2$ could simply
replace its review for $p_2$ with the review for~$p_1$.
This attack represents someone copy-pasting his review for the wrong paper into one of his reviews.

We note that for the given specification of the workflow, such an attack is unavoidable in ``real life'', as it can
be performed also outside the workflow system. Concretely, $a_2$ can
directly communicate the reviews for $p_1$ to $a_1$ through any
communication channel, for instance by~email. 
To combat this attack, the example should be changed to having disjunct reviewing groups --- 
whenever a reviewer $r$ is assigned to a paper $p$, no one else that has a conflict with the other assigned papers of $r$ can be assigned to $p$.
\end{example}


\subsection{Verification}
\label{sec:verif}

As hinted in~\cref{sec:hyperltl}, given a non-omitting workflow $w$
and an $\exists^*_\pi\forall^*_\pi\bsfotl$ formula $\psi$ denoting a
set of bad behaviors, our approach for checking whether
$w\models\neg\psi$ consists in checking the (un)satisfiability of the
formula $\psi'$ given in the proof of
Theorem~\ref{thm:workflows_hltl_sat}(1).

As an instance of this approach, we obtain that Non-interference with
Declassification can be checked on non-omitting workflows. 

\begin{theorem}\label{thm:noninterference}
  For any non-omitting workflow, it is decidable to check whether it
  satisfies Non-interference with Declassification for a finite number
  of \textcausal agents and an unbounded number of \textstubborn agents, as
  long as for each formula~$\phi$ expressing a declassification
  condition, the negation normal form of $\neg\phi$ contains no existential
  quantifier.
\end{theorem}

It is easy to check that the negation of non-interference can be brought
into $\exists^*_\pi\forall^*_\pi\bsfotl$. Then, as $w$ is
non-omitting, the result follows directly from by
\cref{thm:workflows_hltl_sat}(2).

In \cite{ATVA16}, the authors show that for workflows without loops, 
it is possible to check non-interference even when \emph{all} agents behave in a \textcausal way.
This is no longer the case for workflows with loops:

\begin{theorem}
    The problem of checking for a given non-omitting workflow $w$ 
    wether it satisfies Non-interference with Declassification for an \emph{unbounded} number of \textcausal agents
    is undecidable, even if for all formulas~$\phi$ expressing a declassification
    condition, the negation normal form of $\neg\phi$ contains no existential
    quantifier.
\end{theorem}

\begin{proof}
    As in the proof for \cref{thm:omitting_undecidable}, we present a reduction from the periodic tiling problem.
    We will consider a workflow $w$ over signature $\sig$ with
    \[
    \sig = \big( \{A\}, \{ \afirst \}, \{ Q,O,\obs,\adjy,T'_0,\ldots,T'_{k-1} \}, \ar \big)
    \]
    where $A$, $\afirst$, $Q$, and $T'_0,\ldots,T'_{k-1}$ are as in proof for \cref{thm:omitting_undecidable}, and they fulfill the same purposes. 
    The \adjy symbols denote again a vertical adjacency relation, but here it is not filled with input data, but rather computed stepwise by the workflow. 
    The relation denoted by $O$ and $\obs$ contain an initial secret that differs on both traces and spreads along the adjacency relation \adjy.
    The symbols $Q,T'_0,\ldots,T'_{k-1}$ denote again high-input relations containing input data with a declassification of $\ttrue$ (i.e. they are always equal in both traces).

    We consider the following workflow:
    \begin{align*}
    &\wforall\ .\ \Oracle(\afirst) \to \obs \wadd (\afirst)\\
    &\wloop\\
    &\qquad        \text{\% Information flow from a to b}\\
    &\qquad        \wforall\ a, b\ \wmay.\ \ttrue \to \adjy \wadd (b, a)\\
    &\qquad        \text{\% Clear \adjy}\\
    &\qquad        \wforall\ a,b.\ \adjy \wsub (a,b)
    \end{align*}

    We add the rest of the tiling requirements to the declassification condition of $O$, so that there only is an information flow violation in case that all formulas hold.

    As we again use time as the $x$-axis, we reuse \cref{eq:omitting_assign1,eq:omitting_assign2,eq:omitting_Q1,eq:omitting_Q2,eq:omitting_lrcomp,eq:omitting_lrperiod}.
    We also use $\alast$ as one representative agent of the bottom-most row, so we reuse \cref{eq:omitting_tbperiod} to specify that $\alast$ is compatible to $\afirst$.
    This time $\alast$ is not part of the signature, but we will call the outermost agent of the non-interference condition $\alast$, so all declassification conditions can use the variable.

    \noindent Two adjacent agents need to be assigned \ycomp\ tiles (\cref{eq:causal_tbcomp}).
    \begin{equation}\label{eq:causal_tbcomp}
        \forall a, b.\ (\LTLfinally \adjy(b,a)) \implies \LTLglobally \bigvee_{\ycomp(i,j)} T'_i(a) \land T'_j(b)
    \end{equation}

    \noindent The last agent should be reachable from the first via $\adjy$ relations.
    This is expressed by specifying that $\alast$ can observe different tuples on traces $\pi,\pi'$ (the non-interference property.).
    Let $\psi$ be the conjunction of equations \cref{eq:omitting_assign1,eq:omitting_assign2,eq:omitting_Q1,eq:omitting_Q2,eq:omitting_lrcomp,eq:omitting_lrperiod,eq:causal_tbcomp}.
    Let the declassification conditions be:
    \[
      \phi_{\Oracle} = \neg \psi,\ \phi_{Q} = \ttrue,\ \phi_{T'_i} = \ttrue\\
    \]
    We then verify the non-interference property 
    \begin{equation}\label{eq:causaltbreach}
    \begin{multlined}
      \forall \pi, \pi', \alast. \forall a \neq \alast.\ \causal_{\pi,\pi'}(a)\\
      \implies \noninter(\pi,\pi',\alast)
    \end{multlined}
    \end{equation}

    There exists a satisfying model for the negation of the property in \cref{eq:causaltbreach} on $w$ iff there is a periodic tiling for $(T,\xcomp,\ycomp)$.
    If $\alast$ observes different low outputs (tuples in $\adjy$), either he is the same agent as $\afirst$ and $y$-compatible to himself or there exists a chain of \textcausal agents spreading the tuples along $\adjy$ to $\alast$. 
    Since $\afirst$ is the only one able to read $\Oracle$, every possible differences can only originate in $\obs$.
    Thus, there is a chain of $y$-compatible \textcausal agents $a_i$ starting with $\afirst$ that reaches $\alast$.
    We can construct the tiling from any satisfying model in exactly the same way as in the proof of \cref{thm:omitting_undecidable}.
\end{proof}

\section{Experimental Evaluation}
\label{sec:evaluation}

We have implemented our approach into the tool \niwo.%
\footnote{The source code together with all examples can be found on the authors' website.}
Our tool takes as input the specification of a workflow together with declassification conditions
and the number of \textcausal agents.
From that, it generates a sorted \fotl formula whose satisfiability is equivalent
to the existence of a violation of the non-interference property.
For non-omitting workflows, this formula is further compiled into an equi-satisfiable
\ltl formula to be checked by some of-the-shelf \ltl satisfiability solver.
Currently, we use \aalta~\cite{aalta} for that purpose.
The \textsc{NIWO} tool is, to our knowledge, the first implementation of an automated verification approach for workflows.

\subsection{Size of the formulas}

We consider the structure and size of the formulas whose unsatisfiability is checked in order to establish whether non-interference holds for a given workflow. 
Such a formula is a conjunction with four conjuncts. 
One conjunct is a universally quantified formula that describes the semantics of the workflow (see \cref{sec:verif}).
The next conjunct represents the agent model and it is an existentially quantified formula with the number of quantifiers matching the given upper bound on the number of \textcausal agents.
The third conjunct represent the assumption on the control flow and it is a propositional formula. 
The last conjunct describes the existence of a counterexample to the non-interference property and it is an existentially quantified formula.
When considering only \textstubborn agents, for a workflow with observable relations of maximum arity $n_s$ per sort $s$, the formula to be checked for unsatisfiability uses $\sum_s n_s$ existential quantifiers.
Every existentially quantified variable adds one Skolem constant to the smallest universe (for its sort) that can be considered.
As described in the proof of \cref{thm:decidability} universal quantification over a sort is translated into 
a conjunction over all Skolem constants of that sort.\footnote{Note that, nevertheless, sorts may greatly reduce the number of conjuncts in comparison to the unsorted case.}
Thus, the resulting encoding of the universally quantified conjunct is exponential in the number of existentially 
quantified variables of each sort.
For every \textcausal agent considered, $\sum_s n_s$ additional existential quantifiers are added to the formula.
Since every \textcausal agent adds multiple existential quantifiers, the resulting \ltl formula can be orders of magnitude 
larger when considering multiple \textcausal agents of the same sort.

\subsection{Experiments}

We evaluated \niwo on realistic example workflows, among these the example from the introduction,
as well as synthetic examples to evaluate scalability issues.

We used several realistic examples.
\emph{Notebook} is an event-based model of a notebook-like data structure where several people can write messages, but everyone can only read his own data.
It is proven safe by our implementation, even in the presence of \textcausal agents.
\emph{Conference} is the example conference management from \cref{ex:easychair},
where our implementation finds the counterexample described in \cref{sec:verif}.
\emph{Conference-acceptance} is a slight variation that forgoes reviews and replaces it by an acceptance relation.
Since it is very similar to the initial conference example, we use it to showcase the impact of small changes to the workflow to the verification problem.
\emph{Conference-linear} is the motivating example used in \cite{ATVA16}.
It is a simpler version of the Conference example, which does not use loops, and exhibits a very similar attack.
\emph{University} is an example from a university environment where a professor writes down secret grading information for students. It takes at least 2 conspiring agents for a student to learn something about grades of other students.

Additionally, we used synthetic examples to illustrate the scalability of the approach in several dimensions.
The \emph{Fixed-Arity-X} examples show the behavior when increasing the number of relations.
These cases contain $X$ relations that are successive copies of each other starting from some secret input.
The \emph{Fixed-Arity-X-safe} examples are similar, but are devoid of counterexamples.
The \emph{Sorted-Increasing-Arity-X} and \emph{Increasing-Arity-X} examples show the impact of using sorts.
They contain $X$ relations of arities $1, \ldots, X$, respectively. For every relation of arity $n$, a tuple containing the first $n-1$ variables has to be present in the relation with arity $n - 1$.
For the \emph{Increasing-Arity} cases, all variables refer to the same sort, whereas in the \emph{Sorted-Increasing-Arity} variant, every relation of arity $n$ uses $n$ different sorts.
\emph{\textcausal-X} and \emph{Sorted-\textcausal-X} cases showcase the scalability with the number of \textcausal agents that are part of the attack. These cases are set up in a way that a successful attack needs to consist of at least $X$ \textcausal agents.

\ifinv
\else  
\begin{table*}
\caption{\label{fig:data}Experiment Results}
\begin{tabular}{| l | r | r || r | r | r | r | r |}
\hline
Name & \# \textcausal agents & Workflow size & Result & Universe size & \fotl size & \ltl size & Time (s)\\
\hline
\hline
Notebook & 0 & 2 & safe & (1,1) & 266 & 240 & 0.18\\
Notebook & 1 & 2 & safe & (3,2) & 309 & 993 & 2.92\\
\hline
Conference & 0 & 5 & safe & (2,1,1) & 628 & 1089 & 2.50\\
Conference & 1 & 5 & unsafe & (4,2,2) & 700 & 8771 & 91.86 \\
\hline
Conference-acceptance & 0 & 5 & safe & (2,1) & 628 & 1089 & 2.47\\
Conference-acceptance & 1 & 5 & unsafe & (4,2) & 700 & 5187 & 45.63\\
\hline
Conference-linear & 0 & 4 & safe & (2,1) & 469 & 698 & 0.75\\
Conference-linear & 1 & 4 & unsafe & (4,2,1) & 541 & 4116 & 4.91\\
\hline
University & 0 & 3 & safe & (1,1) & 305 & 202 & 0.01 \\
University & 2 & 3 & unsafe & (4,3,3,2) & 408 & 2727 & 1.28 \\
\hline
\hline
Fixed-Arity-10 & 0 & 10 & unsafe & (2) & 1928 & 5299 & 0.89\\
Fixed-Arity-15 & 0 & 15 & unsafe & (2) & 3963 & 11114 & 3.09\\
Fixed-Arity-20 & 0 & 20 & unsafe & (2) & 6723 & 19054 & 16.85 \\
Fixed-Arity-10-safe & 0 & 10 & safe & (2) & 1924 & 5283 & 33.40 \\
Fixed-Arity-15-safe & 0 & 15 & safe & (2) & 3959 & 11098 & 158.83 \\
Fixed-Arity-20-safe & 0 & 20 & safe & (2) & 6719 & 19038 & 740.91 \\
\hline
Increasing-Arity-2 & 0 & 2 & safe & (2) & 180 & 335 & 0.08 \\
Increasing-Arity-3 & 0 & 3 & safe & (3) & 301 & 2206 & 6.20 \\
Increasing-Arity-4 & 0 & 4 & safe & (4) & 451 & 21894 & - \\
Sorted-Increasing-Arity-2 & 0 & 2 & safe & (1,1) & 180 & 163 & 0.03 \\
Sorted-Increasing-Arity-3 & 0 & 3 & safe & (1,1,1) & 301 & 270 & 0.09\\
Sorted-Increasing-Arity-5 & 0 & 5 & safe & (1,1,1,1,1) & 630 & 559 & 0.40\\
Sorted-Increasing-Arity-10 & 0 & 10 & safe & (1,\ldots,1) & 1960 & 1719 & 8.97 \\
\hline
\Textcausal-1 & 0 & 4 & safe & (2) & 654 & 1129 & 8.23 \\
\Textcausal-1 & 1 & 4 & unsafe & (4) & 747 & 2805 & 6.05\\ 
\Textcausal-2 & 0 & 6 & safe & (2) & 778 & 1353 & 26.73 \\
\Textcausal-2 & 2 & 6 & unsafe & (6) & 965 & 6338 & 195.31 \\
Sorted-\Textcausal-2 & 2 & 3 & unsafe & (4,3,1) & 378 & 1184 & 1.47 \\ 
Sorted-\Textcausal-3 & 3 & 4 & unsafe & (5,4,4,1) & 598 & 3169 & 1.96 \\ 
Sorted-\Textcausal-5 & 5 & 5 & unsafe & (7,6,6,6,6,1) & 1197 & 14510 & 17.05 \\
\hline
\end{tabular}
\end{table*}
\fi

\subsection{Results}
The results of the experiments are shown in \cref{fig:data}.
The first column describes the number of \textcausal agents that are considered. 
All other participating agents are considered as \textstubborn as per \cref{sec:prop}.

The size of the workflow is the number of blocks the workflow consists of, not counting choice and loop constructs.
The result is \emph{safe} iff the \ltl formula was proven \emph{unsatisfiable} by \aalta and \emph{unsafe} otherwise.
The next column gives the sizes of the considered universes.
For example, to show that \emph{Conference} is safe with respect to one \textcausal agent, it is enough to consider universes containing $4$ reviewers, $2$ papers and $2$ reviews (one per paper), respectively. The universes' sizes are given as a tuple, for instance $(4,2,2)$.  
The size of both the \fotl and \ltl formulas is the number of nodes in the formulas abstract syntax tree. 
The last column is the time (in seconds) that it takes \aalta to check the satisfiability of the \ltl formula (averaged over $10$ runs).
All experiments were carried out on a desktop machine using an Intel i7-3820 clocked at $3.60$ GHz with $15.7$ GiB of RAM and running Debian with a timeout of 20 minutes.

The implementation is able to handle all examples based on real applications in less than $100$ seconds.
Even though the size of the resulting formula is exponential in the number of agents in the universe,
\aalta was still able to check the satisfiability of formulas consisting of thousands of \ltl operators in reasonable time.
As expected of a satisfiability solver, giving a counterexample for a formula is almost always faster than proving it unsatisfiable for formulas of comparable complexities.

The \emph{Fixed-Arity} cases show that workflows handle an increasing number of relations with the same arity quite well.
Here, adding another relation increases the size of the formula by a small factor, since the size of the needed universe stays the same - only the universally quantified encoding of the control flow graph grows.
Increasing the necessary arity of the relations increases the minimum size of the universe - as shown by the \emph{Increasing-Arity} cases.
In case that all necessary agents are of the same sort, the formula grows exponentially, whereas it grows a lot slower in case that increasing the arity introduces a new sort.
Since in those cases the size of the needed universe is exactly one agent per sort, the resulting \ltl formula is even smaller than the \fotl specification. 
The biggest factor in increasing the state space of the workflow, however, is the number of necessary \textcausal agents as shown by the two variants of the \emph{\Textcausal-X} cases.
Since every \textcausal agent that we consider adds another copy of all of the agents needed to verify the workflow for only \textstubborn agents, adding the first \textcausal agent doubles the minimum amount of agents in the universe. Since the size of the LTL formula is exponential in the number of agents, adding more \textcausal agents causes the size of the resulting \ltl formula to grow rapidly.

\section{Related Work}

The closest work to ours is \cite{ATVA16} where a similar workflow language is introduced. 
That language, however, does not provide control-flow constructs such as loops.
Accordingly, a bounded model checking approach suffices to verify hyperproperties
such as non-interference.  
In presence of loops, bounded model checking does no longer suffice for that purpose.

The workflow model that we consider is a type of a multi-agent system.
Another type of multi-agent systems is represented by business
processes. They are often described by BPMN
diagrams~\cite{dijkman2008semantics} and formalized by Petri nets.
A business process is a collection of activities, and a workflow
thereof represents the flow of data items between activities.
Activities are performed by users, who may need to synchronize on
certain actions, but otherwise execute activities asynchronously.
This is contrast to our formalism, where workflow steps are executed
synchronously by a set of agents.
Information flow in business processes has been considered, e.g.,
in~\cite{BauereissH14}.  There the MASK framework for possibilistic
information flow security~\cite{MASK00}, and in particular a variant
of the unwinding technique from~\cite{GoguenMeseguer82}, is used to
prove that specifications satisfying particular constraints are safe.
Up to our understanding, the approach is not easily amenable to
automation.

There have recently been many efforts to verify concrete workflow
systems, such as conference management
systems~\cite{Kanav2014,Arapinis2012} or an eHealth
system~\cite{ehealth}, or a social media platform~\cite{CoSMeDis17}.
For instance, the \textsc{CoCon} conference management
system~\cite{Kanav2014} is implemented and checked in the interactive
theorem prover \textsc{Isabelle}.
Its security model uses a specialized non-interference notion (based on
nondeducibility~\cite{sutherland1986model}), which is motivated by the
need for fine-grained declassification conditions. In our case, this
need is satisfied by the use of \fotl, which allows for specifying
fine-grained declassification conditions both in the ``what'' and in
the ``when''
directions~\cite{DimensionsAndPrinciplesOfDeclassification}.
ConfiChair~\cite{Arapinis2012} is a cryptographic-based model of a
cloud-based conference management system for which the strong secrecy
(also a hyperproperty) of paper contents and reviews is automatically
checked with the ProVerif tool. 
In contrast to these works, which focus on the verification of one
specific system, we propose a modeling language for workflow, together
with a verification approach.

Another attempt to verify parametric systems via a formalization in first-order logic is
the CSDN language~\cite{Sagiv_SDN}. That language has been proposed for describing and 
verifying the semantics of controllers in software-defined networks (SDNs). 
With our workflow language, it shares that the semantics is specified in
terms of relations.
A CSDN program consists of a sequence of controller rules, each guarded by an event pattern. 
When the event pattern fires, the corresponding command is executed.
Commands are expressed in a simple imperative
language, which allows to query and update relations.
In contrast, the basic step of our workflow language is the \wforall
block, which consists of a sequence of guarded updates to relations.
This sequence is executed \emph{in parallel} for each instantiation of
the block variables.
Accordingly, the semantics in~\cite{Sagiv_SDN} and the one in the present
paper are orthogonal, and cannot easily simulate one another.
Furthermore, in contrast to~\cite{Sagiv_SDN}, we are not only
interested in plain invariants, but temporal non-interference
properties expressed by \hfotl.

Expressing trace properties with sorted \fotl has been initiated
already in the work of Manna and Pnueli~\cite{MannaPnueli81} and
logic-based approaches are now standard in the verification of such
properties.
Logic-based approaches for non-trace properties are less common and
include the ones based on
epistemic temporal logics~\cite{ReasoningAboutKnowledgeBook},
\mbox{SecLTL}~\cite{SecLTL}, and in particular
HyperLTL~\cite{clarkson2014temporal}, the logic whose first-order
extension we use in this paper.


\section{Conclusion}

We have provided an extension to the workflow language from~\cite{ATVA16} with non-deterministic
control-flow structures. We have encoded the semantics of these workflows as well
as complex non-interference properties into sorted \fotl
and identified a fragment of sorted \fotl where satisfiability is decidable. 
From that, we concluded that non-interference is decidable for non-omitting  workflows
and a fixed number of \textcausal agents.
These methods are strong enough to automatically construct attacks to the example property.

We also explored in how far our decidability result can be further generalized.
We found, however, that dropping either the restriction on workflows or the bound on the number
of \textcausal agents results in undecidability.

We have implemented the tool \niwo which automatically verifies 
sorted non-interference properties for non-omitting workflows.
We evaluated our implementation on workflows inspired by a conference management system.
Nonetheless, we would like to see specifications of larger workflows in order to better 
understand the potentials and limitations of our methods.

A practical verification system for arbitrary workflows in our language 
requires to deal with the satisfiability problem of general (sorted) \fotl formulas.
Clearly, \fotl is a fragment of~\fol --- using one unary function symbol.
It remains for future work to explore in how far current automated theorem provers
such as 
\textsc{Spass}~\cite{weidenbach2009spass} or 
Z3~\cite{de2008z3}, or model-finders such as 
\textsc{Alloy}~\cite{jackson2012software} or its temporal extension Electrum~\cite{Electrum}
are able to deal with the non-interference formulas for workflows;
or what extra proving technology is required.

\begin{acks}
We thank the anonymous reviewers for their feedback. 
This work was partially supported by the German Research Foundation (DFG)
under the project ``SpAGAT'' (grant no.~FI 936/2-1) in the priority program ``Reliably Secure Software Systems – RS3'', in the doctorate program ``Program and Model Analysis - PUMA'' (no. 1480), and as part of the Collaborative Research Center ``Methods and Tools for Understanding and Controlling Privacy'' (SFB 1223).
\end{acks}

\bibliographystyle{ACM-Reference-Format}
\bibliography{refs}

\end{document}